\documentclass[letterpaper, 10 pt, conference]{ieeeconf}
\usepackage[utf8]{inputenc}

\IEEEoverridecommandlockouts

\usepackage{balance}

\usepackage{multirow}

\usepackage{algorithm}
\usepackage{tcolorbox}

\usepackage{amsmath,mathtools}  
\usepackage{amsfonts,amssymb}
\usepackage{bbm}
\usepackage{xspace}    

\usepackage{graphicx}
\graphicspath{{figs/}}

\usepackage{subcaption}
\usepackage{caption}

\usepackage{hyperref}
\hypersetup{colorlinks=true, linkcolor= blue, allcolors=blue, citecolor = blue, filecolor=black, urlcolor=blue}

\usepackage{todonotes}
\usepackage{comment}

\usepackage{tabularx}

\usepackage{diagbox}
\newcolumntype{K}[1]{>{\centering\arraybackslash}p{#1}}

\makeatletter
\newcommand{\multiline}[1]{%
  \begin{tabularx}{\dimexpr\linewidth-\ALG@thistlm}[t]{@{}X@{}}
    #1
  \end{tabularx}
}
\makeatother

\usepackage[per-mode=symbol]{siunitx}

\usepackage{cite}  
\usepackage{array}  
\usepackage{booktabs}           

\makeatletter
\def\th@plain{%
  \itshape 
}
\def\@begintheorem#1#2{\trivlist
   \item[\hskip\labelsep\bfseries #1\ #2.]} 
\def\@opargbegintheorem#1#2#3{\trivlist
   \item[\hskip\labelsep\bfseries #1\ #2\ (#3).]} 
\makeatother

\newtheorem{theorem}{Theorem}

\newcounter{remark}
{\par\endtrivlist\unskip}

\newenvironment{dpsolution}{%
  \par\vspace{3pt}\noindent\textbf{DP Solution:}\quad
}{\par\endtrivlist\unskip}

\newcounter{problem}
\newenvironment{problem}{%
\par\vspace{3pt}\noindent\refstepcounter{problem}\textbf{Problem~\theproblem:}}%
{\par\endtrivlist\unskip}

\usepackage[extramath,probability,reviewmark]{mylatexdefs}

\usepackage{stfloats}

\usepackage{hyperref}
\usepackage{graphicx}
\usepackage{amsmath,mathtools} 
\usepackage{mathrsfs}
\usepackage{amsfonts,amssymb}
\usepackage{upgreek}
\usepackage{bbm}
\usepackage{dsfont}
\usepackage{algorithm}
\usepackage{algpseudocode}

\usepackage{tikz}
\usetikzlibrary{arrows,positioning,plotmarks,arrows.meta}

\usepackage{pgfplots}
\usepgfplotslibrary{patchplots,groupplots}
\usepackage{grffile}

\pgfplotsset{compat=newest}  

\pgfplotsset{
  every axis/.append style={
    width=\columnwidth,
    height=0.7\columnwidth,
    label style={font=\large},
    ticklabel style={font=\normalsize},
    title style={font=\Large},
    legend style={font=\normalsize}
  }
}
\usepackage{mathrsfs}
\usetikzlibrary{arrows}

\newcommand{\betaclc}{\ensuremath{\beta}}
\newcommand{\Real}{\ensuremath{\mathbb{R}}}

\newcommand{\realA}{\ensuremath{\hat{A}}}
\newcommand{\realB}{\ensuremath{\hat{B}}}
\newcommand{\modelA}{\ensuremath{A}}
\newcommand{\modelB}{\ensuremath{B}}
\newcommand{\xonerealp}{\ensuremath{\hat{x}_{1}}}
\newcommand{\xtworealp}{\ensuremath{\hat{x}_{2}}}

\newcommand{\xtplusonerealp}{\ensuremath{\hat{x}_{t+1}}}
\newcommand{\xrealp}{\ensuremath{\hat{x}_{1:T}}}
\newcommand{\Jr}{\ensuremath{J_{\mathrm{r}}}}

\newcommand{\Jc}{\ensuremath{J_{\mathrm{c}}}}

\newcommand{\xonerealspace}{\ensuremath{\mathscr{\hat{X}}_1}}

\newcommand{\xtrealspace}{\ensuremath{\mathscr{\hat{X}}_t}}
\newcommand{\xTrealspace}{\ensuremath{\mathscr{\hat{X}}_T}}
\newcommand{\xrealspace}{\ensuremath{\mathscr{\hat{X}}}}
\newcommand{\utrealspace}{\ensuremath{\mathscr{U}_t}}
\newcommand{\urealspace}{\ensuremath{\mathscr{U}}}
\newcommand{\xtmodelspace}{\ensuremath{\mathscr{X}_t}}
\newcommand{\xmodelspace}{\ensuremath{\mathscr{X}}}
\newcommand{\clcpolicy}{\ensuremath{\mathbf{g}^{\mathrm{clc}}}}
\newcommand{\optpolicy}{\ensuremath{\mathbf{g}^{\mathrm{*}}}}
\newcommand{\normal}{\ensuremath{\mathcal{N}}}

\begin{document}

\title{\LARGE \bf Combined Learning and Control: A New Paradigm for Optimal Control with Unknown Dynamics}
\author{Panagiotis Kounatidis$^{2}$, {\IEEEmembership{Student Member, IEEE}}, and Andreas A. Malikopoulos$^{2,3}$, {\IEEEmembership{Senior Member, IEEE}}
\thanks{This research was supported in part by NSF under Grants CNS-2149520, CMMI-2348381, IIS-2415478, and in part by Mathworks.}
\thanks{$^{2}$Systems Engineering Program, Cornell University, Ithaca, NY 14850 USA.}
\thanks{$^{3}$Systems Engineering Program and School of Civil and Environmental Engineering, Cornell University, Ithaca, NY 14853 USA.}
\thanks{Emails: {\tt\small \{pk586,amaliko\}@cornell.edu}.}
}

\maketitle

\begin{abstract}
In this paper, we present the combined learning-and-control (CLC) approach, which is a new way to solve optimal control problems with unknown dynamics by unifying model-based control and data-driven learning. The key idea is simple: we design a controller to be optimal for a proxy objective built on an available model while penalizing mismatches with the real system, so that the resulting controller is also optimal for the actual system. Building on the original CLC formulation, we demonstrate the framework to the linear–quadratic regulator problem and make three advances: (i) we show that the CLC penalty is a sequence of stage-specific weights rather than a single constant; (ii) we identify when these weights can be set in advance and when they must depend on the (unknown) dynamics; and (iii) we develop a lightweight learning loop that tunes the weights directly from data without abandoning the benefits of a model-based design. We provide a complete algorithm and an empirical study against common baseline methods. The results clarify where prior knowledge suffices and where learning is essential, and they position CLC as a practical, theoretically grounded bridge between classical optimal control and modern learning methods.
\end{abstract}


\section{Introduction}
Optimal control concerns synthesizing a sequence of inputs that steer a dynamical system while minimizing a prescribed performance criterion. When a reliable model is available, classical approaches—the calculus of variations, Pontryagin’s minimum principle, and dynamic programming—provide systematic characterizations of optimal policies and practical numerical methods \cite{kirk2004,bertsekas2017}. In many modern applications, however, the dynamics are uncertain or only partially known, and model mismatch can degrade performance, motivating approaches that blend model-based structure with data-driven learning.

\subsection{Model-based control}
The calculus of variations formulates the optimal control problem as a functional optimization task. Its fundamental theorem establishes necessary conditions for a trajectory to be optimal, namely that the first variation of the cost functional vanishes along optimal trajectories. In general, enforcing this condition leads to a nonlinear two-point boundary value problem that typically lacks closed-form solutions. Numerical methods, such as gradient descent, can be employed to solve this boundary value problem and obtain open-loop optimal controls \cite{kirk2004}. In the special case of linear system dynamics with quadratic cost functionals (LQR), the necessary conditions simplify to a first-order matrix differential equation of the Riccati type. When integrated backward in time, this Riccati equation yields the optimal control law, which takes the form of a time-varying linear state-feedback controller. Pontryagin’s minimum principle extends the calculus of variations by incorporating state and control constraints into the optimization. While it provides general necessary conditions for optimality, its application in practice is often heuristic and tailored to the specific problem structure \cite{bertsekas2017}.

Dynamic programming (DP) formulates the optimal control problem as a sequential, multi-stage Markov decision process \cite{bertsekas2017}. The resulting optimal control law follows from the principle of optimality, which states that at any stage, the minimum cost-to-go equals the sum of the immediate transition cost and the minimum cost-to-go from the subsequent stage onward. In this way, the original functional optimization problem is reduced to a parameter optimization problem with respect to the control inputs. An important feature of the DP framework is that state and control constraints can be incorporated naturally. For certain classes of problems, such as the linear–quadratic regulator (LQR), the DP recursion admits closed-form solutions for the optimal control law. In general, however, DP must be implemented numerically, which requires discretization of the state and control spaces \cite{bertsekas2017}. For high-dimensional problems, this discretization leads to prohibitive computational and memory requirements \cite{larson1967}.

All of the aforementioned approaches to optimal control require full knowledge of the system dynamics. A common way to circumvent this requirement is to assume a model of the dynamics and then apply the same methodologies using the model. While straightforward, this approach often results in suboptimal strategies due to discrepancies between the assumed model and the true system.
\subsection{Learning-based control}
Reinforcement learning (RL), in contrast, enables optimal control without prior knowledge of the system’s dynamics. Most RL algorithms rely on stochastic approximation of the Bellman equation to estimate the cost-to-go function \cite{bertsekas1996}. Another major class of RL methods, known as policy search, optimizes the parameters of a stochastic control policy directly via stochastic gradient descent \cite{sutton1999, recht2019}. Modern RL approaches typically integrate function approximation of the cost-to-go with policy search and deep neural networks \cite{sutton2020}, thereby enhancing scalability. Comprehensive surveys of RL algorithms can be found in \cite{recht2019, kiumarsi2018}.
A key limitation of RL is its reliance on multiple trajectories (episodes) to learn the optimal policy. In contrast, adaptive control \cite{ioannou2006} seeks to identify or adapt the control law online, using only a single trajectory. For example, \cite{bradtke1994} demonstrated how the Q-function of the LQR problem can be learned online via recursive least squares. Another approach, iterative learning control  \cite{armstrong2021}, also aims at online performance improvement but requires the system to be repeatedly reset to the same initial state, thereby mimicking the episodic structure of RL \cite{zhang2019}. Successful applications of learning-based control in autonomous vehicles include learning-based multi-robot navigation \cite{le2025}, autonomous racing \cite{wischnewski2020}, \cite{spielberg2023}, \cite{rosolia2018}, traffic control \cite{wu2022}, \cite{jang2019} and real-time learning of powertrain systems with respect to the driver's driving style \cite{malikopoulos2008}.

An alternative approach to the optimal control problem with unknown dynamics is the combined learning-and-control (CLC) framework. The theoretical foundations of this approach were first developed for general classes of systems in \cite{Malikopoulos2022a} and later specialized to linear systems in \cite{Malikopoulos2024}. CLC derives a control strategy by minimizing a proxy cost function that depends only on a nominal model of the system. This proxy cost is parameterized by two elements: (i) a parameter $\beta$, which steers the resulting strategy toward the true optimal control law, and (ii) all possible real state trajectories, which ensure that the strategy remains consistent with the actual system dynamics. Consequently, the strategy produced by CLC is guaranteed to satisfy the real dynamics and be optimal with respect to the proxy cost and —crucially—with respect to the original cost functional. For this equivalence to hold, however, $\beta$ must be appropriately chosen. To date, the CLC framework \cite{Malikopoulos2022a,Malikopoulos2024} has not addressed how to select the parameter $\beta$, nor whether this selection can be made \emph{a priori}, independently of the true system dynamics.

\subsection{Contributions}
In this paper, we analyze and extend the CLC framework in several key directions. First, we demonstrate that $\beta$ is a parameter vector, with one component for each decision stage of the optimal control problem. Second, we establish theoretical results that characterize the boundary of the system class for which $\beta$ can be selected a priori, that is, independently of the true system dynamics. Beyond this boundary, the optimal choice of $\beta$ necessarily depends on the real dynamics. Motivated by this observation, we augment the CLC algorithm with a learning framework that enables the online identification of the optimal $\beta$ values, thereby preserving the effectiveness of the CLC methodology. Finally, we present the complete CLC algorithm, integrated with the learning framework, and evaluate its performance on the LQR problem with unknown dynamics, comparing it against benchmark reinforcement learning algorithms. The code of this paper is publicly available at \href{https://github.com/Panos20102k/Learning-LQR}{https://github.com/Panos20102k/Learning-LQR}.

\subsection{Organization}
The remainder of the paper is organized as follows. In Section~\ref{sec:CLC}, we present the CLC framework in its general form, as originally developed in \cite{Malikopoulos2022a, Malikopoulos2024}. In Section~\ref{sec:prb}, we introduce the class of systems considered in this study—scalar, linear, time-invariant systems with quadratic cost functions. This restriction enables us to precisely identify the boundary of the system class for which $\beta$ can be chosen a priori, without dependence on the true system dynamics. We also briefly discuss existing approaches to the LQR problem with unknown dynamics. In Section~\ref{sec:algo}, we describe the implementation of the CLC algorithm. In Section~\ref{sec:theory}, we present theoretical results that delineate the conditions under which $\beta$ can be selected independently of the real dynamics, and when it cannot. In Section~\ref{sec:learning}, we introduce a learning framework that resolves this dependence and preserves the effectiveness of CLC. In Section~\ref{sec:results}, we apply the proposed algorithm to the LQR problem with unknown dynamics and compare its performance with benchmark reinforcement learning methods. Finally, in Section~\ref{sec:conclusions}, we provide concluding remarks and discuss directions for future research.




\section{Combined Learning and Control (CLC)}
\label{sec:CLC}

In this section, we review the CLC framework \cite{Malikopoulos2022a,Malikopoulos2024}.
We consider a real system together with an available nominal model of its dynamics. Let $X_t \in \mathbb{R}^n$, $n \in \mathbb{N}$, denote the model state at time $t$, and let $\hat{X}_t \in \mathbb{R}^n$ denote the state of the real system. The control input is $U_t \in \mathbb{R}^m$, $m \in \mathbb{N}$, the disturbance is $W_t \in \mathbb{R}^r$, $r \in \mathbb{N}$, and the measurement noise is $Z_t \in \mathbb{R}^s$, $s \in \mathbb{N}$. The model dynamics evolve as
\begin{align}
X_{t+1} = A_t X_t + B_t U_t + D_t W_t, \quad t=0,\ldots,T-1, 
\label{eq:model}
\end{align}
while the real system evolves as
\begin{align}
\hat{X}_{t+1} = \hat{A}_t \hat{X}_t + \hat{B}_t U_t + \hat{D}_t W_t, \quad t=0,\ldots,T-1.
\label{eq:actual}
\end{align}
Here, $A_t \in \mathbb{R}^{n \times n}$, $B_t \in \mathbb{R}^{n \times m}$, and $D_t \in \mathbb{R}^{n \times r}$ are known matrices, whereas $\hat{A}_t \in \mathbb{R}^{n \times n}$, $\hat{B}_t \in \mathbb{R}^{n \times m}$, and $\hat{D}_t \in \mathbb{R}^{n \times r}$ are unknown. 

At each time $t$, we observe
\begin{align*}
Y_t = C_t X_t + E_t Z_t, \qquad 
\hat{Y}_t = \hat{C}_t \hat{X}_t + \hat{E}_t Z_t,
\end{align*}
where $C_t,\hat{C}_t \in \mathbb{R}^{p \times n}$ and $E_t,\hat{E}_t \in \mathbb{R}^{p \times s}$, with $p \in \mathbb{N}$. 

A control strategy is a sequence $\mathbf{g} = \{g_t; t=0,\ldots,T-1\}$ with
\begin{align}
U_t = g_t(Y_{0:t},U_{0:t-1}),
\label{eq:policy}
\end{align}
where $Y_{0:t} = (Y_0,\ldots,Y_t)$ and $U_{0:t-1} = (U_0,\ldots,U_{t-1})$. Let $\mathcal{G}$ denote the set of admissible strategies. The objective for the actual system is to minimize the total expected cost
\begin{align*}
J(\mathbf{g}) = \mathbb{E}_{\mathbf{g}}\!\Bigg[\sum_{t=0}^{T-1} c_t(\hat{X}_t,U_t) + c_T(\hat{X}_T)\Bigg],
\end{align*}
with stage costs $c_t:\mathbb{R}^n\times\mathbb{R}^m\to\mathbb{R}$ and terminal cost $c_T:\mathbb{R}^n\to\mathbb{R}$. 
\begin{problem}\label{problem:real-general}
    The problem is to find
    \begin{align*}
    \mathbf{g}^* \in \operatorname*{arg\,min}_{\mathbf{g}\in\mathcal{G}} J(\mathbf{g}).
    \end{align*}
\end{problem}
Since the actual matrices in \eqref{eq:actual} are unknown, Problem~\ref{problem:real-general} cannot be solved directly.

To circumvent the lack of knowledge of the actual dynamics, we compress the growing data into a time-invariant sufficient statistic. At time $t$, the information state is defined as
\begin{align}
\Pi_t(X_t,\hat{X}_t) = p(X_t,\hat{X}_t \mid Y_{0:t},U_{0:t-1}).
\label{eq:pi}
\end{align}
The information state is a function of the past observations and controls and serves as a sufficient statistic of the history for optimal decision making. The key structural property is that its evolution does not depend on the particular choice of control strategy but only on its realized action \cite{Malikopoulos2022a}. It has been shown \cite{Malikopoulos2024} that there exists a measurable mapping $\phi_t$ such that
\begin{align}
\Pi_{t+1} = \phi_t(\Pi_t,Y_{t+1},U_t),
\label{eq:piupdate}
\end{align}
which establishes a Markov recursion on a time-invariant space. The passage from \eqref{eq:policy} to \eqref{eq:pi}--\eqref{eq:piupdate} replaces the growing history $(Y_{0:t},U_{0:t-1})$ with the fixed-dimensional object $\Pi_t$; consequently, all subsequent design may be carried out with $\Pi_t$ as the state. In view of \eqref{eq:piupdate}, we restrict attention, without loss of optimality, to separated strategies
\begin{align*}
U_t = g_t(\Pi_t), \qquad \mathbf{g}\in\mathcal{G}_s \subseteq \mathcal{G},
\end{align*}
where the influence of past data on decisions is mediated exclusively through the information state. This separation formalizes the intuition that estimation (updating $\Pi_t$) and control (selecting $U_t$) can be derived independently: the evolution of $\Pi_t$ is unaffected by the particular control law as long as the realized $U_t$ is fed back.

Since the actual system matrices are unknown, we cannot solve Problem~\ref{problem:real-general} offline. Instead, we solve an equivalent offline problem with respect to the known model \eqref{eq:model} and a penalty that aligns the model and actual trajectories in mean square. For a parameter $\beta\in\mathbb{R}$ and a sequence $\hat{x}_{0:T}\in(\mathbb{R}^n)^{T+1}$ representing the expected actual trajectory, define
\begin{align*}
J(\mathbf{g};\hat{x}_{0:T}) 
{}={} & \mathbb{E}_{\mathbf{g}}\!\Bigg[\sum_{t=0}^{T-1} \Big(c_t(X_t,U_t) &+ \beta \,\|X_{t+1}-\hat{x}_{t+1}\|^2\Big) \nonumber\\
& {+}\: c_T(X_T)\Bigg],
\end{align*}
and consider the offline optimization problem.
\begin{problem}\label{problem:clc-general}
Find
    \begin{align}
    \mathbf{g}^\star \in \operatorname*{arg\,min}_{\mathbf{g}\in\mathcal{G}_s} J(\mathbf{g};\hat{x}_{0:T}).
    \end{align}
\end{problem}
Problem~\ref{problem:clc-general} is solved by dynamic programming on the information-state space induced by \eqref{eq:model} and \eqref{eq:piupdate}, yielding an optimal separated law $\mathbf{g}^\star=\{g_t^\star\}$ that is parameterized by $\hat{x}_{0:T}$. Online, we operate the model and the actual system in parallel under $\mathbf{g}^\star$ while computing $\Pi_t$ recursively via \eqref{eq:piupdate}.
\begin{theorem}\label{theorem:basis}
    \cite{Malikopoulos2024} \textit{Let $\mathbf{g}^\star \in \mathcal{G}_s$ denote an optimal separated strategy that solves Problem~\ref{problem:clc-general}. Assume that, during online implementation, the information state $\{\Pi_t\}_{t=0}^T$ defined in \eqref{eq:pi} is available at each time $t$ and evolves recursively according to \eqref{eq:piupdate}. Then the strategy $\mathbf{g}^\star$ is also optimal for Problem~\ref{problem:real-general}, i.e.,
    \begin{align*}
    J(\mathbf{g}^\star) = \inf_{\mathbf{g}\in\mathcal{G}} J(\mathbf{g}).
    \end{align*}}
\end{theorem}
\textit{Proof}. See \cite{Malikopoulos2024}. \hfill $\square$

\section{Problem Formulation}
\label{sec:prb}
The CLC framework introduced above requires the selection of an appropriate value of the parameter $\betaclc$. Only then does the minimization of the proxy cost function yield a control strategy that is also optimal with respect to the original cost functional, as established in Theorem~\ref{theorem:basis}. In this paper, we derive theoretical results that characterize how $\betaclc$ should be chosen for the class of systems under consideration. To this end, we focus on the simplest setting: scalar systems with linear, time-invariant dynamics, classical information structure \cite{Malikopoulos2021}, and quadratic cost functions. For this class, we show that $\betaclc$ can be determined a priori when no penalty is imposed on the control input. When such a penalty is present, however, $\betaclc$ becomes a vector, $\betaclc = (\beta_1, \ldots, \beta_T)$, and all but the final element depend on the true system dynamics, and therefore cannot be determined a priori. To address this limitation, we extend the CLC framework with a learning scheme that estimates the optimal values of $\betaclc$, thereby preserving the effectiveness of the approach.

To this end, we consider the following setup. The evolution of the real system is 
\begin{align}\label{eq:real-system-dyn}
    \hat{X}_{t+1} = \realA \hat{X}_t + \realB U_t, \quad t=0,\ldots,T-1,
\end{align}
where $\realA, \realB \in \Real$, while that of the model is
\begin{align}\label{eq:model-dyn}
    X_{t+1} = \modelA X_t + \modelB U_t, \quad t=0,\ldots,T-1,
\end{align}
where $\modelA, \modelB \in \Real$. The initial state is common to both and given by $\hat{X}_0 = X_0$. The problem we want to solve is
\begin{problem}\label{problem:real}
Find $\optpolicy \in \operatorname*{arg\,min}_{\mathbf{g}\in\mathcal{G}} \Jr(\mathbf{g})$, where
\begin{equation}\label{eq:real-cost}
    \Jr(\mathbf{g}) = \sum_{t=0}^{T-1}\!\big[Q_t \hat{X}_t^2 + R_t U_t^2\big] + Q_T \hat{X}_T^2,
\end{equation}
subject to \eqref{eq:real-system-dyn}, with unknown $\realA$ and $\realB$.
\end{problem}
If $\realA$ and $\realB$ were known, then $\optpolicy = K_t \hat{X}_t$, where $K_t \in \Real$ is a linear, time\text{-}varying state-feedback gain.

\subsection{Existing Methods}
To leverage the linear structure of the optimal policy, many approaches directly learn the state-feedback gains $K_t$ from samples of \Jr\ \cite{fazel2018,li2025}. In practice, these methods often restrict attention to a time-invariant gain $K$ for tractability. For example, policy gradient (PG) posits a linear policy parameterized by a constant gain $K$ and updates $K$ via stochastic gradient descent to reduce \Jr. Although the optimal law is generally time-varying ($K_t$), a constant gain can be a reasonable approximation for time-invariant systems over sufficiently long horizons \cite{bertsekas2017}. To evaluate the current policy, PG injects zero-mean Gaussian exploration into the control input,
\begin{align}\label{eq:pg-explore}
    U_t = K\,X_t + \sigma\,~\eta_t, 
    \eta_t \sim \normal(0,1), 
    \sigma > 0, 
    t=0,\ldots,T-1,
\end{align}
and uses the resulting trajectory cost to form an estimate of $\nabla_{K}\Jr$, thereby enabling stochastic gradient updates of $K$ \cite{yaghmaie2021}.

Random search (RS) \cite{mania2018} is another approach to Problem~\ref{problem:real}. Like PG, it assumes a constant gain $K$ and perturbs it to assess the effect on the cost, but the perturbation is applied directly to $K$ rather than to the control inputs; specifically, $K$ is updated using random directions $\xi \sim \normal(0,1)$ with perturbation magnitude $\sigma \in \Real$.

A different class of methods is Q-learning \cite{watkins1989}. Define the state–action value function for pairs $(\hat{X}_t,U_t)$ by
\begin{align}\label{eq:q-learning}
    Q^{*}(\hat{X}_t,\,U_t) 
    &= c_t + \min_{U_{t+1}} Q^{*}(\hat{X}_{t+1},\,U_{t+1}),
\end{align}
where $\hat{X}_{t+1}$ follows the (unknown) real dynamics, and
\begin{align}\label{eq:stage-cost}
    c_t &=
    \begin{cases}
        Q_t\,\hat{X}_t^2 + R_t\,U_t^2, & t = 0,\ldots, T-1,\\[2pt]
        Q_T\,\hat{X}_T^2, & t = T.
    \end{cases}
\end{align}
A tabular Q-learning update (with discretized state–action spaces) takes the form
\begin{align}\label{eq:q-update}
    Q_{i+1}(\hat{X}_t,\,U_t) 
    &= (1 - \gamma_i)\,Q_i(\hat{X}_t,\,U_t)\\
    &+ \gamma_i\!\left(c_t + \min_{U_{t+1}} Q_i(\hat{X}_{t+1},\,U_{t+1})\right),\\
    & t=0,\ldots,T-1,
\end{align}
and at the terminal stage
\begin{align}\label{eq:q-terminal}
    Q_{i+1}(\hat{X}_T,\,U_T) &= Q_T\,\hat{X}_T^2.
\end{align}
Convergence to $Q^{*}$ is guaranteed provided the stepsizes satisfy
\begin{align}\label{eq:stepsize}
    \sum_{i=0}^{\infty} \gamma_i &= \infty,
    & 
    \sum_{i=0}^{\infty} \gamma_i^2 &< \infty.
\end{align}
A common choice meeting these conditions is \cite{bertsekas2017}: if update $i$ corresponds to the $m$th visit of $(\hat{X}_t,U_t)$, set
\begin{align}\label{eq:stepsize-choice}
    \gamma_i &= \frac{b}{a + m},
    && a,b>0.
\end{align}

\section{Algorithmic Implementation}\label{sec:algo}
In this section, we present how CLC tackles Problem~\ref{problem:real}. Since \Jr\ cannot be evaluated directly (the parameters $\realA$ and $\realB$ are unknown), we minimize a proxy cost \Jc\ that depends only on the model and on parameters $\betaclc = (\beta_1,\ldots,\beta_T)$ and the hypothesized real trajectory $\hat{x}_{1:T}$. Specifically, CLC solves:

\begin{problem}\label{problem:clc}
    Find $\clcpolicy \in \operatorname*{arg\,min}_{\mathbf{g}\in\mathcal{G}} \Jc(\mathbf{g}; \beta, \hat{x}_{1:T})$, where
    \begin{align}\label{eq:proxy-cost}
        \Jc(\mathbf{g}; \beta, \hat{x}_{1:T})
        \;&=\;
        \sum_{t=0}^{T-1}\!\Big(
            Q_t X_t^2 + R_t U_t^2 \nonumber\\
            &+ \beta_{t+1}\,(X_{t+1}-\hat{x}_{t+1})^2
        \Big)
        + Q_T X_T^2.
    \end{align}
\end{problem}
Since $\modelA$ and $\modelB$ are known and $(\beta,\xrealp)$ are fixed, Problem~\ref{problem:clc} can be solved directly.

Let \xtrealspace\ and \xtmodelspace\ denote the spaces of $\hat{X}_t$ and $X_t$, $t=1,\ldots,T$, respectively, and let \utrealspace\ denote the space of $U_t$, $t=0,\ldots,T-1$. Define the product spaces
\[
    \xrealspace \,=\, \textstyle\prod_{t=1}^{T} \xtrealspace,\quad
    \xmodelspace \,=\, \textstyle\prod_{t=1}^{T} \xtmodelspace,\quad
    \urealspace \,=\, \textstyle\prod_{t=0}^{T-1} \utrealspace.
\]
Then a DP solution to Problem~\ref{problem:clc} is:

\begin{dpsolution}
For each $\hat{x}_{1:T}\in\xrealspace = \xonerealspace \times \cdots \times \xTrealspace$, solve the recursion
\begin{align}
    V_T(X_T) \;&=\; Q_T X_T^2, \label{eq:dp-terminal}\\
    V_t(X_t) \;&=\; \min_{U_t\in \utrealspace}
        \Big\{
            Q_t X_t^2 + R_t U_t^2 \nonumber\\
            &+ \beta_{t+1}\,(X_{t+1}-\hat{x}_{t+1})^2
            + V_{t+1}(X_{t+1})
        \Big\}, \nonumber\\ &\quad t=0,\ldots,T-1. \label{eq:dp-recursion}
\end{align}
\end{dpsolution}
This yields the control law $U_t(X_t;\xrealp)$, parameterized by the hypothesized real trajectory $\xrealp\in\xrealspace$. To implement DP, the spaces \xtrealspace, \xtmodelspace, and \utrealspace\ are discretized and finite, and $U_t(X_t;\xrealp)$ is stored as a lookup table.

Next, to enforce the real dynamics, we solve the coupled equations
\begin{align}\label{eq:coupled-system}
    \xtplusonerealp \;=\; \realA\, X_t \;+\; \realB\, U_t\big(X_t;\xrealp\big),
    \qquad t=0,\ldots,T-1.
\end{align}
This system is coupled because each $U_t$ depends on the entire trajectory $\xrealp$. Moreover, $\realA$ and $\realB$ are unknown; thus, black-box root-finding methods are required to solve \eqref{eq:coupled-system}. For small-scale problems (e.g., $T=2$), direct search over the lookup table $U_t(X_t;\xrealp)$ is effective (as used in this paper).

Once a solution $\xrealp^{\mathrm{s}}$ to \eqref{eq:coupled-system} is found, the control strategy applied to the real system,
\[
    \clcpolicy \;=\; \{\,U_t(X_t;\xrealp^{\mathrm{s}})\,\}_{t=0}^{T-1},
\]
is fully determined, since the model dynamics \eqref{eq:model-dyn} and $X_0$ are known. By construction, $\clcpolicy$ simultaneously minimizes \Jc\ and aligns with the real dynamics \eqref{eq:real-system-dyn}. To ensure that $\clcpolicy$ is also optimal for \Jr—our ultimate objective—we select $\betaclc=(\beta_1,\ldots,\beta_T)$ appropriately; then Theorem~\ref{theorem:basis} guarantees that the solution of Problem~\ref{problem:clc} coincides with $\optpolicy$, the solution to Problem~\ref{problem:real}. The procedure is summarized in Algorithm~\ref{alg:clc}.

\begin{algorithm}
\caption{CLC Algorithm}\label{alg:clc}
\begin{algorithmic}[1]   
    \Require \betaclc\,$=(\beta_1,\ldots,\beta_T)$ and $X_0$, \xrealspace\,,\xmodelspace\,,\urealspace\,
    \State Solve Problem~\ref{problem:clc} through DP.
    \State Solve \eqref{eq:coupled-system} for each $t=0,\ldots,T-1$.
    \State Obtain \clcpolicy\,, for which \clcpolicy $=$ \optpolicy\, holds if \betaclc\, was selected appropriately.
\end{algorithmic}
\end{algorithm}

In the next section, we present theoretical results that prescribe the values of \betaclc \(=(\beta_1,\ldots,\beta_T)\) and a learning framework that maintains the effectiveness of the CLC algorithm in cases where \betaclc\ cannot be prescribed a priori.

\section{Theoretical Results}\label{sec:theory}
Let \(\beta^* = (\beta_1^*,\ldots,\beta_T^*)\) denote the optimal \(\beta\)-values, i.e., those for which the policy \clcpolicy\ resulting from Algorithm~\ref{alg:clc} coincides with \optpolicy. We now present results that prescribe the optimal \(\beta\)-values for the CLC algorithm and delineate the boundary of the system class for which this is possible.

\begin{theorem}\label{theorem:no-effort}
\textit{For the class of systems defined by \eqref{eq:real-system-dyn}, \eqref{eq:model-dyn}, and \eqref{eq:real-cost} with \(R_t = 0\) for \(t=0,\ldots,T-1\) and \(\modelB=\realB\), Algorithm~\ref{alg:clc} yields optimal control for \(\beta_t^* = -Q_t + \epsilon\), \(t=1,\ldots,T\), regardless of \(\realA\) and \(\modelA\), as \(\epsilon \to 0\).}
\end{theorem}

\textit{Proof}. We derive \optpolicy\ and the CLC policy \clcpolicy\ and show they match for \(T=1\), \(T=2\), and hence for any finite \(T\).

The optimal control strategy for Theorem~\ref{theorem:no-effort} is
\begin{align}\label{eq:no-effort-opt-policy}
    \optpolicy = \Big\{-\tfrac{\realA}{\realB}X_0,\; -\tfrac{\realA}{\realB}\hat{X}_1,\; \ldots,\; -\tfrac{\realA}{\realB}\hat{X}_{T-1}\Big\}.
\end{align}

\paragraph*{Case \(T=1\)} With \(\Jc = Q_1 X_1^2 + \beta_1 (X_1 - \xonerealp)^2\) and \(X_1=\modelA X_0+\modelB U_0\), minimizing \(\Jc\) gives
\begin{align}\label{eq:no-effort-one-step-u0}
    \frac{\partial \Jc}{\partial U_0} = 0
    \;\Rightarrow\;
    U_0
    = \frac{\beta_1\, \xonerealp}{\modelB\,(Q_1+\beta_1)} - \frac{\modelA}{\modelB} X_0.
\end{align}
The coupling equation
\begin{align}\label{eq:no-effort-one-step-x1hat}
    \xonerealp
    = \realA X_0 + \realB U_0
    = \Bigg[\frac{\realA \modelB - \modelA \realB}{Q_1 \modelB}\Bigg](Q_1+\beta_1) X_0
\end{align}
follows from \eqref{eq:no-effort-one-step-u0}. Substituting \eqref{eq:no-effort-one-step-x1hat} into \eqref{eq:no-effort-one-step-u0} and taking \(\beta_1=-Q_1+\epsilon\) yields
\begin{align}\label{eq:no-effort-one-step-u0-final}
    U_0
    = \Bigg[
        -\frac{\realA}{\modelB}
        + \frac{\modelA \realB}{\modelB^2}
        + \frac{(\realA \modelB - \modelA \realB)\,\epsilon}{Q_1 \modelB^2}
        - \frac{\modelA}{\modelB}
    \Bigg] X_0,
\end{align}
which, since \(\modelB=\realB\) and \(\epsilon\to 0\), gives \(U_0 = -\frac{\realA}{\realB} X_0\), i.e., \(\optpolicy(1)\).

\paragraph*{Case \(T=2\)} The DP for Problem~\ref{problem:clc} (with \(R_t=0\)) is
\begin{align}
    V_2(X_2) &= Q_2 X_2^2 + \beta_2 (X_2 - \xtworealp)^2, \label{eq:no-effort-V2}\\ \nonumber
    V_1(X_1) &= \min_{U_1}\, \big\{ Q_1 X_1^2 + \beta_1 (X_1 - \xonerealp)^2 + V_2(X_2) \big\}\\
               &\doteq \min_{U_1} J_1, \label{eq:no-effort-V1}\\
    V_0(X_0) &= \min_{U_0}\, V_1(X_1) \;\doteq\; \min_{U_0} J_0. \label{eq:no-effort-V0}
\end{align}
Minimizing \(J_1\) gives
\begin{align}\label{eq:no-effort-two-step-u1}
    \frac{\partial J_1}{\partial U_1} = 0
    \;\Rightarrow\;
    U_1 = \frac{\beta_2\, \xtworealp}{\modelB\,(Q_2+\beta_2)} - \frac{\modelA}{\modelB} X_1.
\end{align}
Substituting \eqref{eq:no-effort-two-step-u1} into \(V_1\) and minimizing \(J_0\) yields
\begin{align}\label{eq:no-effort-two-step-u0}
    \frac{\partial J_0}{\partial U_0} = 0
    \;\Rightarrow\;
    U_0 = \frac{\beta_1\, \xonerealp}{\modelB\,(Q_1+\beta_1)} - \frac{\modelA}{\modelB} X_0.
\end{align}
The coupling equations are
\begin{align}
    \xtworealp &= \realA \xonerealp + \realB U_1, \label{eq:no-effort-couple-x2hat}\\
    \xonerealp &= \realA X_0 + \realB U_0. \label{eq:no-effort-couple-x1hat}
\end{align}
With \(\beta_1=-Q_1+\epsilon\), \(\beta_2=-Q_2+\epsilon\), we obtain
\begin{align}
    \xtworealp &= \frac{\realA \modelB \,\xonerealp - \modelA \realB \, X_1}{\modelB \epsilon - \realB(-Q_2+\epsilon)}\,\epsilon, \label{eq:no-effort-two-step-xtworealp}\\
    \xonerealp &= \frac{\realA \modelB - \modelA \realB}{\modelB \epsilon - \realB(-Q_1+\epsilon)}\,\epsilon\, X_0. \label{eq:no-effort-two-step-xonerealp}
\end{align}
Substituting \eqref{eq:no-effort-two-step-xonerealp} into \eqref{eq:no-effort-two-step-u0} and taking \(\modelB=\realB\), \(\epsilon\to 0\) yields \(U_0 = -\frac{\realA}{\realB} X_0 = \optpolicy(1)\). Likewise, substituting \eqref{eq:no-effort-two-step-xtworealp} into \eqref{eq:no-effort-two-step-u1} and using \(X_1=\modelA X_0+\modelB U_0\) gives
\begin{align}\label{eq:no-effort-two-step-u1-final}
    U_1
    = \Bigg[\frac{(-Q_2+\epsilon)\,\modelA}{\modelB Q_2} - \frac{\modelA}{\modelB}\Bigg](\realA-\modelA)X_0
    \;\xrightarrow[\epsilon\to 0]{}\; 0,
\end{align}
which matches \(\optpolicy(2) = -\frac{\realA}{\realB}\hat{X}_1 = 0\). Since \(\optpolicy(t)=0\) for \(t\ge 2\), the result holds for any finite \(T\). \hfill\(\square\)

\begin{theorem}\label{theorem:general}
\textit{For the class of systems defined by \eqref{eq:real-system-dyn}, \eqref{eq:model-dyn}, and \eqref{eq:real-cost} with \(R_t \neq 0\) for \(t=0,\ldots,T-1\) and \(\modelB=\realB\), the optimal value is \(\beta_T^* = -Q_T\). However, \(\beta_t^*\) for \(t=1,\ldots,T-1\) depends on \(\realA\) and therefore cannot be prescribed a priori.}
\end{theorem}

\textit{Proof}. We prove optimality of \(\beta_2=-Q_2\) for \(T=2\); by the principle of optimality this implies \(\beta_T=-Q_T\) for any finite \(T\ge 2\). We then show that \(\beta_t\) depends on \(\realA\) for \(t=1,\ldots,T-1\).

For \(T=2\), the optimal control for Problem~\ref{problem:real} (from DP) is
\begin{align}\label{eq:general-optimal}
    \optpolicy = \Big\{ -\tfrac{Q_2 \realA \realB}{R_1+Q_2 \realB^2}\,\hat{X}_1,\;
                         -\tfrac{P \realA \realB}{R_0+P \realB^2}\,X_0 \Big\},
\end{align}
where
\begin{align}\label{eq:general-P}
    P =
    Q_1
    + R_1\!\left(\frac{Q_2 \realA \realB}{R_1+Q_2 \realB^2}\right)^{\!2}
    + Q_2\!\left(\realA - \frac{Q_2 \realA \realB^2}{R_1+Q_2 \realB^2}\right)^{\!2}\!.
\end{align}
The CLC DP for \(T=2\) is
\begin{align}
    V_2(X_2) &= Q_2 X_2^2 + \beta_2 (X_2 - \xtworealp)^2, \label{eq:general-V2}\\
    V_1(X_1) &= \min_{U_1}\,\big\{ Q_1 X_1^2 + \beta_1 (X_1 - \xonerealp)^2 \nonumber\\ &+ R_1 U_1^2 + V_2(X_2) \big\}
                \;\doteq\; \min_{U_1} J_1, \label{eq:general-V1}\\
    V_0(X_0) &= \min_{U_0}\,\big\{ R_0 U_0^2 + V_1(X_1) \big\}
                \;\doteq\; \min_{U_0} J_0. \label{eq:general-V0}
\end{align}
Minimizing \(J_1\) yields
\begin{align}\label{eq:two-step-u1}
    \frac{\partial J_1}{\partial U_1} = 0
    \;\Rightarrow\;
    U_1
    &= \frac{\beta_2 \modelB\, \xtworealp}{R_1 + \modelB^2 (Q_2+\beta_2)}\nonumber\\
      &- \frac{(Q_2+\beta_2)\,\modelA \modelB}{R_1 + \modelB^2 (Q_2+\beta_2)}\, X_1.
\end{align}
With \(\modelB=\realB\) and \(\beta_2=-Q_2\), the coupling equation
\begin{align}\label{eq:two-step-xtworealp}
    \xtworealp
    = \frac{R_1 \realA}{R_1 + Q_2 \realB^2}\, \xonerealp,
\end{align}
substituted into \eqref{eq:two-step-u1} gives
\begin{align}\label{eq:two-step-u1-final}
    U_1 = -\frac{Q_2 \realA \realB}{R_1 + Q_2 \realB^2}\, \xonerealp,
\end{align}
which equals \(\optpolicy(1)\) since CLC enforces \(\xonerealp=\hat{X}_1\) in Algorithm~\ref{alg:clc}. By the principle of optimality, this establishes \(\beta_T=-Q_T\) for any finite \(T\ge 2\).

Substituting \eqref{eq:two-step-u1-final} into \(V_1\) and minimizing \(J_0\) gives
\begin{align}\label{eq:two-step-u0}
    U_0
    = -\frac{Q_2 \modelA \modelB\, \xtworealp - \beta_1 \modelB\, \xonerealp
              + \modelA \modelB (Q_1+\beta_1) X_0}
             {R_0 + \modelB^2 (Q_1+\beta_1)}.
\end{align}
Using \eqref{eq:two-step-xtworealp} and \(\modelB=\realB\) yields
\begin{align}\label{eq:two-step-u0-final}
    U_0
    = &- \frac{Q_2 \modelA \realB \realA R_1 - \beta_1 \realB (R_1 + Q_2 \realB^2)}
             {(R_1 + Q_2 \realB^2)\,[\,R_0 + \realB^2 (Q_1+\beta_1)\,]}\, \xonerealp
      \;\nonumber\\ &-\; \frac{\modelA \realB (Q_1+\beta_1)}{R_0 + \realB^2 (Q_1+\beta_1)}\, X_0.
\end{align}
Solving \(\xonerealp = \realA X_0 + \realB U_0\) with \eqref{eq:two-step-u0-final} produces a value of \(U_0\) that equals \(\optpolicy(2)\) when \(\beta_1=\beta_1^*(\realA)\); thus \(\beta_1^*\) depends on \(\realA\). Hence, by the principle of optimality, \(\beta_t^*\) depends on \(\realA\) for all \(t=1,\ldots,T-1\). \hfill\(\square\)

\begin{figure}[th]
    \centering
    \resizebox{\columnwidth}{!}{\input{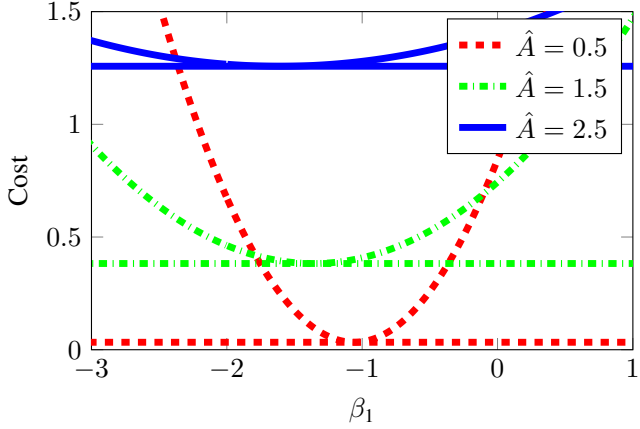}}
    \caption{Optimal $\beta_1$ dependence on $\realA$.}
    \label{fig:dependence}
\end{figure}

The horizontal lines represent the optimal cost for each \realA\ instance. The quadratic curves show the costs achieved by CLC, namely $\Jr(\clcpolicy)$, for various $\beta$ values. We observe that different values of \realA\ yield different optimal $\beta_1^*$ such that $\Jr(\clcpolicy)=\Jr(\optpolicy)$. Hence, for $T=2$, $\beta_1^*$ depends on \realA. By the principle of optimality, the optimal decision at stage $t$ depends on the optimal cost-to-go from stage $t{+}1$; therefore, $\beta_t$ for $t=1,\ldots,T{-}2$ also depends on \realA. \hfill $\square$

The implication of Theorem~\ref{theorem:general} is that the optimal $\beta_t^*$, $t=1,\ldots,T-1$, cannot be determined prior to applying the CLC algorithm. Moreover, the optimal values $\beta_t^*$ are not identical across stages $t=1,\ldots,T$, the nature of which in earlier expositions of CLC \cite{Malikopoulos2022a,Malikopoulos2024} was not investigated.

\section{Learning Framework}\label{sec:learning}
We extend CLC with a learning framework that estimates the optimal $\beta$ values, thereby preserving its effectiveness (at the expense of additional computation, compared later with RL baselines). We first present the algorithm and then establish convergence under standard conditions.
\vspace*{0.05cm} 
\begin{algorithm}[H]
\caption{Learning $\beta^*$ Algorithm}\label{alg:learning}
\begin{algorithmic}[1]
    \State For the current $\beta=(\beta_1,\ldots,\beta_T)$, run Algorithm~\ref{alg:clc}.
    \State Obtain $\clcpolicy(\beta)$, apply it to the real system, and compute $\Jr(\clcpolicy(\beta))$.
    \State Estimate the gradient $\nabla_\beta \Jr(\beta)$.
    \State Update $\beta \leftarrow \beta - \alpha_k \nabla_\beta \Jr(\beta)$ (with stepsize $\alpha_k>0$).
\end{algorithmic}
\end{algorithm}
\begin{theorem}\label{theorem:learning}
\textit{Suppose the composite objective $\widetilde{J}(\beta)\coloneqq \Jr(\clcpolicy(\beta))$ is convex and has a Lipschitz-continuous gradient on the feasible set. Then Algorithm~\ref{alg:learning} converges to $\beta^*=(\beta_1^*,\ldots,\beta_T^*)$ for which $\clcpolicy(\beta^*)=\optpolicy$.}
\end{theorem}

\textit{Proof}. The mapping $\beta \mapsto \clcpolicy(\beta)$ (through Algorithm~\ref{alg:clc}) induces the composite loss $\widetilde{J}(\beta)=\Jr(\clcpolicy(\beta))$. Under the stated assumptions, gradient descent with a suitable stepsize rule converges to a minimizer of $\widetilde{J}$; at $\beta^*$, the induced policy equals \optpolicy. \hfill $\square$

Algorithm~\ref{alg:learning} requires $\nabla_\beta \Jr(\beta)$, but (i) \Jr\ is unknown a priori, and (ii) the map $\beta \mapsto \clcpolicy(\beta)$ is not available in closed form. A practical estimate uses forward finite differences on the composite objective:
\begin{align}
    \nabla_\beta \Jr(\beta)
    &= \Bigg[\frac{\partial \Jr(\beta)}{\partial \beta_1},\ \ldots,\ \frac{\partial \Jr(\beta)}{\partial \beta_T}\Bigg]^{\mathrm{T}}, \label{eq:grad-def}\\
    \frac{\partial \Jr(\beta)}{\partial \beta_t}
    &\approx \frac{\Jr\big(\clcpolicy(\beta+\delta e_t)\big)-\Jr\big(\clcpolicy(\beta)\big)}{\delta},\nonumber\\
    &\qquad t=1,\ldots,T, \ \ \delta>0, \label{eq:fd}
\end{align}
where $e_t$ is the $t$th canonical basis vector. Such finite-difference schemes are theoretically justified with robustness guarantees in related LQR settings \cite{li2025}.

\section{Simulation Results}\label{sec:results}
\begin{figure}[th]
    \centering
    \resizebox{\columnwidth}{!}{
%
%
\begin{tikzpicture}

\begin{axis}[%
  width=0.8\columnwidth,
  height=0.5\columnwidth,
  scale only axis,
  xmin=1, xmax=21,
  xlabel={Iterations},
  xlabel style={font=\large, color=black},
  ymin=-1.6, ymax=2,
  ylabel={$\beta_{1}$},
  ylabel style={font=\large, color=black},
  ticklabel style={font=\normalsize},
  title style={font=\Large},
  legend style={
    legend cell align=left,
    align=left,
    draw=black,
    font=\normalsize
  },
  axis background/.style={fill=white},
  axis lines=box
]
\addplot [color=red, dashed, line width=2.5pt, mark=o, mark options={solid, red}]
  table[row sep=crcr]{%
1	2\\
2	1.3775\\
3	0.852499999999999\\
4	0.592499999999998\\
5	0.217499999999999\\
6	-0.110000000000001\\
7	-0.365000000000002\\
8	-0.560000000000002\\
9	-0.725000000000002\\
10	-0.860000000000002\\
11	-0.957500000000002\\
12	-1.0625\\
13	-1.1375\\
14	-1.1825\\
15	-1.275\\
16	-1.32\\
17	-1.2975\\
18	-1.3\\
19	-1.3025\\
20	-1.305\\
21	-1.28\\
};
\addlegendentry{$\beta_1$}

\addplot [color=red, line width=2.5pt]
  table[row sep=crcr]{%
1	-1.5\\
2	-1.5\\
3	-1.5\\
4	-1.5\\
5	-1.5\\
6	-1.5\\
7	-1.5\\
8	-1.5\\
9	-1.5\\
10	-1.5\\
11	-1.5\\
12	-1.5\\
13	-1.5\\
14	-1.5\\
15	-1.5\\
16	-1.5\\
17	-1.5\\
18	-1.5\\
19	-1.5\\
20	-1.5\\
21	-1.5\\
};
\addlegendentry{$\beta_1^{*}$}

\end{axis}
\end{tikzpicture}%
}
    \caption{Learning $\beta^{*}$.}
    \label{fig:learning}
\end{figure}
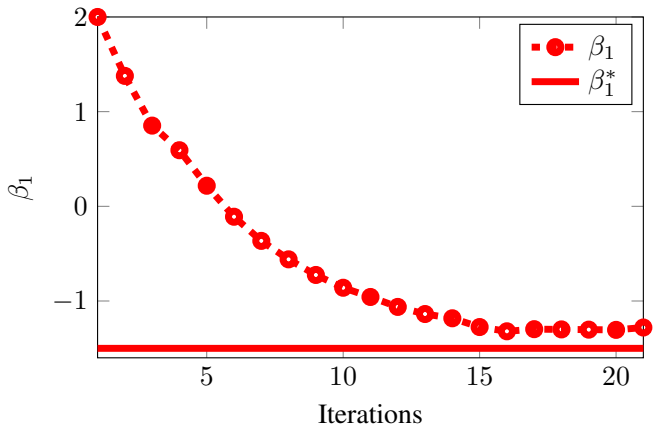

In this section, we apply the CLC algorithm and compare its performance with benchmark RL methods. The real system is given by \eqref{eq:real-system-dyn} with $\realA=2$, $\realB=1$, and $X_0=0.5$, while the model is given by \eqref{eq:model-dyn} with $\modelA=1$ and $\modelB=1$. The cost parameters are $Q_0=0$, $Q_t=1$ for $t\in\{1,2\}$, and $R_t=1$ for $t\in\{0,1\}$. The CLC algorithm requires selecting $\beta=(\beta_1,\beta_2)$. As a consequence of Theorem~\ref{theorem:general}, we set $\beta_2=-Q_2$, whereas $\beta_1$ must be learned. For this small-scale instance, we solve Problem~\ref{problem:clc} via closed-form dynamic programming (as in the proof of Theorem~\ref{theorem:general}) and obtain the optimal value $\beta_1^*=-1.5$, i.e., $\Jr\!\big(\clcpolicy(\beta^*=(-1.5,-Q_2))\big)=\Jr(\optpolicy)$. Figure~\ref{fig:learning} illustrates the convergence of Algorithm~\ref{alg:learning} to $\beta_1^*=-1.5$; the $x$\text{-}axis reports the iterations of Algorithm~\ref{alg:learning}.

\begin{figure}[th]
    \centering
    \resizebox{\columnwidth}{!}{\input{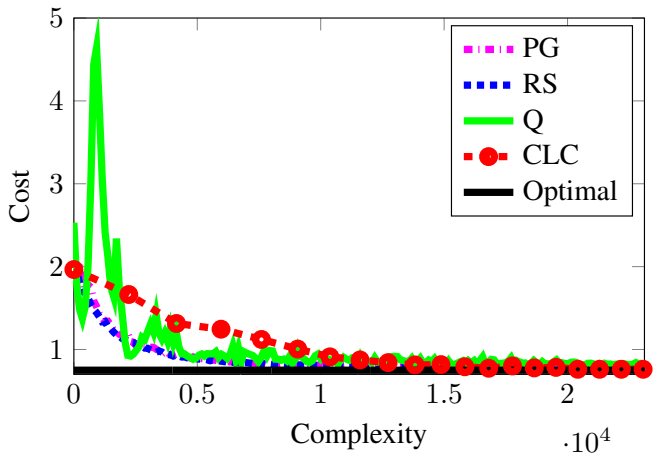}}
    \caption{Comparison with reinforcement learning algorithms.}
    \label{fig:comparison}
\end{figure}

Next, we evaluate the RL baselines introduced in Section~\ref{sec:prb}—policy gradient (PG), random search (RS), and Q-learning (Q)—on the same problem instance. Figure~\ref{fig:comparison} reports the comparison in terms of sample efficiency: the $x$\text{-}axis shows the number of real-system trajectories (episodes) generated by each method, and the $y$\text{-}axis shows the resulting cost $\Jr$ of the synthesized control policy at that sample budget (lower is better).

We observe that CLC and Q-learning require more samples than PG and RS to approach the optimal policy. This is because PG and RS \emph{assume} a linear state-feedback structure and learn its parameters (Section~\ref{sec:prb}); they therefore begin with the correct inductive bias for this instance. By contrast, CLC and Q-learning make no such structural assumption and thus exploit fewer problem-specific simplifications. This lack of bias, however, makes them suitable for more general optimal control problems, including those with nonlinear optimal feedback laws. In particular, CLC equipped with its learning framework can, in principle, learn any $\beta$ that minimizes \Jr. The caveat is that \Jr\ may be nonconvex, in which case convergence to a unique global optimum is not guaranteed.

Regarding computational complexity, CLC generates real-system trajectories only in Step~2 of Algorithm~\ref{alg:clc}, when solving the coupled equations \eqref{eq:coupled-system}. In our experiment, we use direct search over the lookup table $U_t(X_t;\xrealp)$ to solve \eqref{eq:coupled-system}, which is effective for small instances like the one considered here. For larger problems, efficiency can be improved by employing more sophisticated black-box root-finding methods that (i) do not require knowledge of the real dynamics and (ii) can handle coupled fixed-point equations. Consequently, the overall complexity of CLC can be further reduced as \eqref{eq:coupled-system} is solved more efficiently.

\section{Conclusions}\label{sec:conclusions}
We presented the CLC approach for the LQR problem with unknown dynamics. We derived conditions for selecting the parameter vector $\beta=(\beta_1,\ldots,\beta_T)$, showing when $\beta$ can be chosen \emph{a priori} and when it must be learned due to dependence on the true dynamics. For the latter case, we introduced a learning framework that estimates $\beta$ and preserves the efficacy of CLC. We evaluated CLC on an LQR instance and compared it against reinforcement learning baselines. As expected, PG and RS---which assume a linear state-feedback structure---exhibited superior sample efficiency on this linear task, whereas CLC and Q-learning, which make fewer structural assumptions, were less sample efficient but more broadly applicable. Notably, the CLC\,+\,learning framework can, in principle, discover any $\beta$ that minimizes the original cost $\Jr$, enabling nonlinear optimal policies when present. Finally, we noted that CLC’s computational burden is dominated by solving the coupled equations in \eqref{eq:coupled-system}; more efficient black-box solvers can further reduce this cost.

A potential direction for future research includes extending CLC to settings with multiple controllers operating under nonclassical information structures, where agents have heterogeneous and asymmetric observations and may signal through their control actions. 

The code of this paper is publicly available at \href{https://github.com/Panos20102k/Learning-LQR}{https://github.com/Panos20102k/Learning-LQR}.

\balance

\bibliographystyle{IEEEtran}
\bibliography{IEEEabrv,refs,refs_IDS,literature_panos}

\begin{thebibliography}{10}
\providecommand{\url}[1]{#1}
\csname url@rmstyle\endcsname
\providecommand{\newblock}{\relax}
\providecommand{\bibinfo}[2]{#2}
\providecommand\BIBentrySTDinterwordspacing{\spaceskip=0pt\relax}
\providecommand\BIBentryALTinterwordstretchfactor{4}
\providecommand\BIBentryALTinterwordspacing{\spaceskip=\fontdimen2\font plus
\BIBentryALTinterwordstretchfactor\fontdimen3\font minus
  \fontdimen4\font\relax}
\providecommand\BIBforeignlanguage[2]{{%
\expandafter\ifx\csname l@#1\endcsname\relax
\typeout{** WARNING: IEEEtran.bst: No hyphenation pattern has been}%
\typeout{** loaded for the language `#1'. Using the pattern for}%
\typeout{** the default language instead.}%
\else
\language=\csname l@#1\endcsname
\fi
#2}}

\bibitem{kirk2004}
D.~Kirk, \emph{Optimal Control Theory: An Introduction}.\hskip 1em plus 0.5em
  minus 0.4em\relax Dover Publications, 2004.

\bibitem{bertsekas2017}
D.~P. Bertsekas, \emph{Dynamic Programming and Optimal Control}, 4th~ed.\hskip
  1em plus 0.5em minus 0.4em\relax Athena Scientific, 2017.

\bibitem{larson1967}
R.~Larson, ``A survey of dynamic programming computational procedures,''
  \emph{IEEE Transactions on Automatic Control}, vol.~12, no.~6, pp. 767--774,
  1967.

\bibitem{bertsekas1996}
D.~P. Bertsekas and J.~N. Tsitsiklis, \emph{Neuro-Dynamic Programming}.\hskip
  1em plus 0.5em minus 0.4em\relax Athena Scientific, 1996.

\bibitem{sutton1999}
\BIBentryALTinterwordspacing
R.~S. Sutton, D.~McAllester, S.~Singh, and Y.~Mansour, ``Policy gradient
  methods for reinforcement learning with function approximation,'' in
  \emph{Advances in Neural Information Processing Systems}, S.~Solla, T.~Leen,
  and K.~M\"{u}ller, Eds., vol.~12.\hskip 1em plus 0.5em minus 0.4em\relax MIT
  Press, 1999. [Online]. Available:
  \url{https://proceedings.neurips.cc/paper_files/paper/1999/file/464d828b85b0bed98e80ade0a5c43b0f-Paper.pdf}
\BIBentrySTDinterwordspacing

\bibitem{recht2019}
\BIBentryALTinterwordspacing
B.~Recht, ``A tour of reinforcement learning: The view from continuous
  control,'' \emph{Annual Review of Control, Robotics, and Autonomous Systems},
  vol.~2, pp. 253--279, May. 2019. [Online]. Available:
  \url{https://doi.org/10.1146/annurev-control-053018-023825}
\BIBentrySTDinterwordspacing

\bibitem{sutton2020}
R.~Sutton and A.~Barto, \emph{Reinforcement Learning: An Introduction},
  2nd~ed.\hskip 1em plus 0.5em minus 0.4em\relax Bradford Books, 2020.

\bibitem{kiumarsi2018}
B.~Kiumarsi, K.~G. Vamvoudakis, H.~Modares, and F.~L. Lewis, ``Optimal and
  autonomous control using reinforcement learning: A survey,'' \emph{IEEE
  Transactions on Neural Networks and Learning Systems}, vol.~29, no.~6, pp.
  2042--2062, 2018.

\bibitem{ioannou2006}
\BIBentryALTinterwordspacing
P.~Ioannou and B.~Fidan, \emph{Adaptive Control Tutorial}.\hskip 1em plus 0.5em
  minus 0.4em\relax Philadelphia, PA: Society for Industrial and Applied
  Mathematics, 2006. [Online]. Available:
  \url{https://epubs.siam.org/doi/abs/10.1137/1.9780898718652}
\BIBentrySTDinterwordspacing

\bibitem{bradtke1994}
S.~Bradtke, B.~Ydstie, and A.~Barto, ``Adaptive linear quadratic control using
  policy iteration,'' in \emph{Proceedings of 1994 American Control Conference
  - ACC '94}, vol.~3, 1994, pp. 3475--3479 vol.3.

\bibitem{armstrong2021}
A.~A. Armstrong, A.~J. Wagoner~Johnson, and A.~G. Alleyne, ``An improved
  approach to iterative learning control for uncertain systems,'' \emph{IEEE
  Transactions on Control Systems Technology}, vol.~29, no.~2, pp. 546--555,
  2021.

\bibitem{zhang2019}
\BIBentryALTinterwordspacing
Y.~Zhang, B.~Chu, and Z.~Shu, ``A preliminary study on the relationship between
  iterative learning control and reinforcement learning,''
  \emph{IFAC-PapersOnLine}, vol.~52, no.~29, pp. 314--319, 2019, 13th IFAC
  Workshop on Adaptive and Learning Control Systems ALCOS 2019. [Online].
  Available:
  \url{https://www.sciencedirect.com/science/article/pii/S2405896319326187}
\BIBentrySTDinterwordspacing

\bibitem{le2025}
\BIBentryALTinterwordspacing
V.-A. Le, P.~Kounatidis, and A.~A. Malikopoulos, ``Combining graph attention
  networks and distributed optimization for multi-robot mixed-integer convex
  programming,'' 2025. [Online]. Available:
  \url{https://arxiv.org/abs/2503.21548}
\BIBentrySTDinterwordspacing

\bibitem{wischnewski2020}
A.~Wischnewski, J.~Betz, and B.~Lohmann, ``Real-time learning of non-gaussian
  uncertainty models for autonomous racing,'' in \emph{2020 59th IEEE
  Conference on Decision and Control (CDC)}, 2020, pp. 609--615.

\bibitem{spielberg2023}
N.~A. Spielberg, M.~Templer, J.~Subosits, and J.~C. Gerdes, ``Learning policies
  for automated racing using vehicle model gradients,'' \emph{IEEE Open Journal
  of Intelligent Transportation Systems}, vol.~4, pp. 130--142, 2023.

\bibitem{rosolia2018}
U.~Rosolia and F.~Borrelli, ``Learning model predictive control for iterative
  tasks. a data-driven control framework,'' \emph{IEEE Transactions on
  Automatic Control}, vol.~63, no.~7, pp. 1883--1896, 2018.

\bibitem{wu2022}
C.~Wu, A.~R. Kreidieh, K.~Parvate, E.~Vinitsky, and A.~M. Bayen, ``Flow: A
  modular learning framework for mixed autonomy traffic,'' \emph{IEEE
  Transactions on Robotics}, vol.~38, no.~2, pp. 1270--1286, 2022.

\bibitem{jang2019}
K.~Jang, E.~Vinitsky, B.~Chalaki, B.~Remer, L.~Beaver, A.~A. Malikopoulos, and
  A.~Bayen, ``Simulation to scaled city: zero-shot policy transfer for traffic
  control via autonomous vehicles,'' in \emph{Proceedings of the 10th ACM/IEEE
  International Conference on Cyber-Physical Systems}, 2019, p. 291–300.

\bibitem{malikopoulos2008}
A.~A. Malikopoulos, D.~N. Assanis, and P.~Y. Papalambros, ``Real-time
  self-learning optimization of diesel engine calibration,'' \emph{Journal of
  Engineering for Gas Turbines and Power}, vol. 131, no.~2, p. 022803, 2008.

\bibitem{Malikopoulos2022a}
A.~A. Malikopoulos, ``Separation of learning and control for cyber-physical
  systems,'' \emph{Automatica}, vol. 151, no. 110912, 2023.

\bibitem{Malikopoulos2024}
------, ``Combining learning and control in linear systems,'' \emph{European
  Journal of Control}, vol.~80, no. Part A, p. 101043, 2024.

\bibitem{Malikopoulos2021}
------, ``On team decision problems with nonclassical information structures,''
  \emph{IEEE Transactions on Automatic Control}, vol.~68, no.~7, pp.
  3915--3930, 2023.

\bibitem{fazel2018}
\BIBentryALTinterwordspacing
M.~Fazel, R.~Ge, S.~M. Kakade, and M.~Mesbahi, ``Global convergence of policy
  gradient methods for the linear quadratic regulator,'' in \emph{International
  Conference on Machine Learning}, 2018. [Online]. Available:
  \url{https://api.semanticscholar.org/CorpusID:51881649}
\BIBentrySTDinterwordspacing

\bibitem{li2025}
\BIBentryALTinterwordspacing
W.~Li, P.~Kounatidis, Z.-P. Jiang, and A.~A. Malikopoulos, ``On the robustness
  of derivative-free methods for linear quadratic regulator,'' 2025. [Online].
  Available: \url{https://arxiv.org/abs/2506.12596}
\BIBentrySTDinterwordspacing

\bibitem{yaghmaie2021}
\BIBentryALTinterwordspacing
A.~Y. Farnaz and L.~Ljung, ``A crash course on reinforcement learning,'' 2021.
  [Online]. Available: \url{https://arxiv.org/abs/2103.04910}
\BIBentrySTDinterwordspacing

\bibitem{mania2018}
\BIBentryALTinterwordspacing
H.~Mania, A.~Guy, and B.~Recht, ``Simple random search of static linear
  policies is competitive for reinforcement learning,'' in \emph{Advances in
  Neural Information Processing Systems}, vol.~31, 2018. [Online]. Available:
  \url{https://proceedings.neurips.cc/paper_files/paper/2018/file/7634ea65a4e6d9041cfd3f7de18e334a-Paper.pdf}
\BIBentrySTDinterwordspacing

\bibitem{watkins1989}
C.~J. C.~H. Watkins, ``Learning from delayed rewards,'' Ph.D. dissertation,
  University of Cambridge, 1989.

\end{thebibliography}

\end{document}